\documentstyle[12pt]{article}
\begin{document}
\thispagestyle{empty}
\begin{raggedleft}
UR-1444\\
hep-th/9511129
\\
Nov/95\\
\end{raggedleft}
$\phantom{x}$\vskip 0.618cm\par
{\huge \begin{center}
ON SOLDERING CHIRALITIES
\end{center}}\par
\vfill
\begin{center}
$\phantom{X}$\\ {\Large R. AMORIM\footnote{Permanent
address: Instituto de F\'\i sica, Universidade Federal do Rio de Janeiro,
Brasil}
\footnote{amorim@urhep.pas.rochester.edu},\,\,
A. DAS and
C. WOTZASEK$\,^1$
\footnote{clovis@urhep.pas.rochester.edu}
 }\\[3ex] {\em Department of Physics and Astronomy\\ University of
Rochester\\ Rochester, NY 14627, USA}
\end{center}
\vfill
\begin{abstract}

\noindent We study how to solder two Siegel chiral bosons into
one  scalar field in a gravitational background.
\end{abstract}
\vfill
\newpage

\section{Introduction}

The research of chiral scalars in two space-time dimensions has attracted
much attention\cite{stone}\cite{string}.  These objects can be obtained
from the restriction of a scalar field to move in one direction only,
as done by Siegel\cite{siegel}, or by a first-order Lagrangian theory,
as proposed by Floreanini and Jackiw\cite{florjack}.  The equivalence of
these two independent descriptions for chiral scalars has been established
both in the context of Senjanovic\cite{senjanovic} formalism of
constrained path-integral in \cite{berson}, and later, in the
operatorial canonical approach
in \cite{kulkulkir}, by gauge-fixing an existing symmetry in Siegel's model.
This procedure will leave behind only one degree of freedom in phase-space,
corresponding to the chiral excitations, just as in the
Floreanini-Jackiw model.

Scalar fields in 2D can be viewed as bosonized versions of
Dirac fermions and chiral bosons can be seen to correspond to
two dimensional versions of Weyl fermions.  In a more formal level, it
has been shown by Sonneschein\cite{sonneschein} and by Tseytlin and
West\cite{tsewest} that, in some sense, the sum of the flat space-time
actions of two chiral scalars of opposite chirality, does correspond
to the action of a single 2D scalar field.  This seems correct since
the number of degrees of freedom adds up
correctly, and a 2D, conformally invariant field theory, is known to
possess two independent current algebras associated with each of the
chiral components.

In a more detailed study, Stone\cite{stone2} has shown that one needs
more than the direct sum of two fermionic representations of the
Kac-Moody group to describe a
Dirac fermion.  Stated differently, the action of a bosonized Dirac
fermion is not simply the sum of the actions of two bosonized Weyl
fermions, or chiral bosons.  Physically, this is connected with the
necessity to abandon the separated right and left symmetries, and
accept that vector gauge symmetry should be preserved at all times.
This restriction will force the two independent chiral scalars to
belong to the same {\it multiplet}, effectively soldering them together.
The basic idea in \cite{stone2} to sew the two de-quantized left and
right Dirac seas of each Weyl component, was to introduce a gauge
field to remove the obstruction to vector gauge invariance.  This gauge
field, being just an auxiliary field, without any dynamics, can
be eliminated in favor of the physically relevant quantities.

In this work we follow the basic physical principle of \cite{stone2}
to solder together two Siegel chiral bosons of opposite chiralities to
establish the formal equivalence with a single scalar field in a
gravitational background. In section 2 we present the gauge procedure necessary
for the soldering of the chiral scalars, obtaining a
Lagrangian that is invariant under vector gauge transformations.
In section 3 we study
the symmetry group of the quoted Lagrangian, showing that it can be
written in a way where the full diffeomorphism invariance
becomes manifest. In section 4 some considerations regarding truncations,
generators of gauge transformations and a hidden duality symmetry are
discussed. Further geometrical considerations are done in section 5. The last
section
is reserved for some final comments and perspectives.

\section{The gauging procedure}

To begin with, lets us review a few known facts about Siegel's
theory\footnote{We are using standard light-front variables:
$x^{\pm}={1\over\sqrt{2}}\left(x^0\pm x^1\right)$}.  First, one
can see this model as the result of gauging
the semi-local affine symmetry\cite{hull}

\begin{equation}
\label{01}
\delta\varphi=\epsilon^-\partial_-\varphi\;\;\; ; \;\;\;\partial_-\epsilon^- =
0
\end{equation}

\noindent possessed by the action of a free scalar field.  This can be done
with the introduction of a gauge field $\lambda_{++}$, as long as it
transforms as

\begin{equation}
\label{02}
\delta \lambda_{++}=-\partial_+\epsilon^- +\epsilon^-\partial_-
\lambda_{++}-\lambda_{++}\partial_-\epsilon^-\,.
\end{equation}

\noindent The result of this procedure is a Siegel action for a left-moving
chiral boson

\begin{equation}
\label{03}
{\cal L}_0^{(+)} = \partial_+\varphi\partial_-\varphi
+ \lambda_{++} \partial_-\varphi\partial_-\varphi\,.
\end{equation}

\noindent One can also interpret this theory as describing the action for
the coupling of scalar field to a chiral ${\cal W}_2$-gravity\cite{hull}.
The gauge field is just a Lagrange multiplier imposing the constraint

\begin{equation}
\label{04}
T_{--}=\partial_-\varphi\,\partial_-\varphi\approx 0\,,
\end{equation}

\noindent known to satisfy the conformal algebra.  Similarly,
one can gauge the semi-local affine symmetry

\begin{equation}
\label{06}
\delta\varphi=\epsilon^+\partial_+\varphi\;\;\; ; \;\;\;\partial_+\epsilon^+ =
0
\end{equation}

\noindent to obtain the right-moving Siegel chiral boson, by introducing
the gauge field $\lambda_{--}$ such that

\begin{equation}
\label{07}
\delta \lambda_{--}=-\partial_-\epsilon^+ +\epsilon^+\partial_+
\lambda_{--}-\lambda_{--}\partial_+\epsilon^+\,.
\end{equation}

In fact, if we write the right and left
chiral boson actions as

\begin{equation}
\label{08}
{\cal L}_0^{(\pm)} = {1\over 2} J_{\pm}(\varphi)\partial_{\mp}\varphi
\end{equation}

\noindent with

\begin{equation}
\label{09}
J_{\pm}(\varphi)=2\left(\partial_{\pm}\varphi +
\lambda_{\pm\pm}\partial_{\mp}\varphi\right)
\end{equation}

\noindent it is easy to verify that these models are indeed invariant
under Siegel's transformations (\ref{01},\ref{02}) and
(\ref{06},\ref{07}), using that

\begin{equation}
\label{010}
\delta J_{\pm}= \epsilon_{\pm}\partial_{\mp}J_{\pm}
\end{equation}

\noindent  It is worth mentioning at this point that Siegel's actions
for left and right chiral boson can be seen as the action for a scalar
field immersed in a gravitational background whose metric is appropriately
truncated.  In this sense, Siegel symmetry for each chirality can be
seen as a truncation of the reparametrization symmetry existing for the
scalar field action.  We should mention that the Noether current $J_+$
defined above is in fact
the non vanishing component of the left chiral current $J_+ = J_{(L)}^-$,
while $J_-$ is the non vanishing component of the right chiral current
$J_- = J_{(R)}^+$, with the left and right currents being defined in terms
of the axial and vector currents as

\begin{eqnarray}
\label{333}
J_\mu^{(L)}=J_\mu^{(A)}+J_\mu^{(V)}\,,\nonumber\\
J_\mu^{(R)}=J_\mu^{(A)}-J_\mu^{(V)}\,.
\end{eqnarray}

Let us next consider the question of the vector gauge symmetry.  We can use
an iterative Noether procedure to gauge the global U(1) symmetry

\begin{eqnarray}
\label{20}
\delta\varphi &=& \alpha\,,\nonumber\\
\delta\lambda_{++}\, &=& 0
\end{eqnarray}

\noindent possessed by Siegel's model (\ref{03}).  Under the action of
the group of transformations (\ref{20}), written now with a local parameter,
the action (\ref{03}) changes as

\begin{equation}
\label{30}
\delta {\cal L}_0^{(+)} = \partial_-\alpha J_+
\end{equation}

\noindent with the Noether current $J_+=J_+(\varphi)$ being
given as in (\ref{09}).
To cancel this piece, we introduce a gauge field $A_-$
coupled to the Noether current, redefining the original Siegel's Lagrangian
density as

\begin{equation}
\label{50}
{\cal L}_0^{(+)}\rightarrow {\cal L}_1^{(+)}={\cal L}_0^{(+)} +A_- J_+
\,,
\end{equation}

\noindent where the variation of the gauge field is defined as

\begin{equation}
\label{60}
\delta A_-=-\partial_-\alpha
\,.
\end{equation}

\noindent As the variation of ${\cal L}_1^{(+)}$ does not
vanish modulo total derivatives, we introduce a further
modification as

\begin{equation}
\label{70}
{\cal L}_1^{(+)}\rightarrow {\cal L}_2^{(+)}={\cal L}_1^{(+)}+
\lambda_{++}A_-^2
\end{equation}

\noindent whose variation gives

\begin{equation}
\label{80}
\delta{\cal L}_2^{(+)} = 2 A_-\partial_+\alpha\,.
\end{equation}

\noindent This left over piece cannot be canceled by a Noether
counter-term, so that a gauge invariant action for $\varphi$
and $A_-$ does not exist, at least with the introduction of only
one gauge field.  We observe, however, that this action has the virtue of
having a variation dependent only on $A_-$ and not on $\varphi$.
Expression (\ref{80}) is a reflection of the standard anomaly
that is intimately connected with the chiral properties of
$\varphi$.

Now, if the same gauging procedure is
followed for an opposite chirality Siegel boson, say

\begin{equation}
\label{90}
{\cal L}_0^{(-)} = \partial_+\rho\partial_-\rho
+ \lambda_{--} \partial_+\rho\partial_+\rho
\end{equation}

\noindent subject to

\begin{eqnarray}
\label{100}
\delta\rho &=& \alpha\,,\nonumber\\
\delta\lambda_{--} &=& 0\,,\nonumber\\
\delta A_+ &=& -\partial_+\alpha\,,
\end{eqnarray}

\noindent then one finds that the sum of the right and left gauged
actions ${\cal L}_2^{(+)}+{\cal L}_2^{(-)}$ can be made gauge
invariant if a contact term of the form

\begin{equation}
\label{110}
{\cal L}_C = 2 A_+ A_-
\end{equation}

\noindent is introduced. One can check that indeed the complete gauged
Lagrangian

\begin{eqnarray}
\label{120}
{\cal L}_{TOT}
 &=& \partial_+\varphi\partial_-\varphi
+ \lambda_{++} \partial_-\varphi\partial_-\varphi +
\partial_+\rho\partial_-\rho
+ \lambda_{--} \partial_+\rho\partial_+\rho +\nonumber\\
&+& A_+J_-(\rho) + A_- J_+(\varphi)+\lambda_{--}A_+^2
+\lambda_{++}A_-^2 +2 A_- \, A_+
\end{eqnarray}

\noindent with $J_{\pm}$ defined in
Eq.(\ref{09}) above, is invariant
under the set of transformations (\ref{20}), (\ref{60}) and
(\ref{100}).  For completeness, we note that Lagrangian
(\ref{120}) can also be written in the form

 \begin{eqnarray}
\label{121}
{\cal L}_{TOT}
 &=& D_+\varphi D_-\varphi
+ \lambda_{++} D_-\varphi D_-\varphi \nonumber\\
&+&D_+\rho D_-\rho
+ \lambda_{--} D_+\rho D_+\rho +\left(\varphi-\rho\right)E\,,
\end{eqnarray}

\noindent modulo total derivatives. In the above expression,
we have introduced the covariant derivatives $D_{\pm}\varphi=
\partial_\pm\varphi+A_\pm$, with a similar expression for $D_\pm\rho$, and
$E\equiv\partial_+A_--\partial_-A_+$.
In form (\ref{121}), ${\cal L}_{TOT}$ is manifestly gauge invariant.

\section{Diffeomorphism}

What may be  the most remarkable consequence of the gauging
procedure we have presented in the previous section is
that the two decoupled Siegel's symmetries, associated with
each sector originally described by the pair
$\varphi\,,\lambda_{++}$ and $\rho\,,\lambda_{--}$, have now been enlarged
to a complete diffeomorphism, while
these quantities have become effectively coupled in a highly non trivial
way  to have full
diffeomorphism invariance. To see how these features occur, we
will first redefine the
fields $A_{\pm}$ by a shift
that would diagonalize the Lagrangian in Eq. ({\ref{120}).
Let

\begin{equation}
\label{130}
\bar A_\pm=  A_\pm-{1\over \Delta}
\left(J_{\pm}-\lambda_{\pm\pm}J_{\mp}\right)\,,
\end{equation}

\noindent where $J_{+}=J_{+}(\varphi)$, $J_{-}=J_{-}(\rho)$ and
$\Delta=2 (\lambda_{++}\lambda_{--}-1)$.
Under this redefinition of the fields, The Lagrangian becomes

\begin{equation}
\label{132}
{\cal L}_{TOT}={\cal L}_g+{\cal L}_{\bar A}
\end{equation}

\noindent where

\begin{eqnarray}
\label{133}
{\cal L}_g
 &=&{1\over 2}
  \sqrt{-g}g^{\alpha\beta}
 \partial_\alpha\Phi\partial_\beta\Phi\,,\nonumber\\
{\cal L}_{\bar A}&=& \lambda_{--}\bar A_+^2
+\lambda_{++}\bar A_-^2 +2 \bar A_- \, \bar A_+\,.
\end{eqnarray}

\noindent  In the above expressions we have introduced the metric tensor
density

\begin{eqnarray}
\label{134}
G^{--}= \sqrt{-g}g^{--}
&=&-4{\lambda_{++}\over\Delta}\,,\nonumber\\
G^{++}= \sqrt{-g}g^{++}&=&-4{\lambda_{--}\over\Delta}\,, \nonumber\\
G^{+-}=\sqrt{-g}g^{+-}&=&-{2\over\Delta}(1+\lambda_{++}\lambda_{--})\,,
\end{eqnarray}

\noindent  as well as the
field

\begin{equation}
\label{135}
\Phi={1\over \sqrt{2}}(\rho-\varphi)\,.
\end{equation}

We observe that in two dimensions  $ \sqrt{-g}g^{\alpha\beta}$
needs only two parameters to be defined in a proper
way. As it should be, $det(\sqrt{-g}g^{\alpha\beta})=-1$.
We also note that because of conformal invariance, we cannot determine
$g_{\alpha\beta}$ itself.

Before  studying the symmetries of the model given by
${\cal L}_{TOT}$, we note that in the path integral
approach, the fields $\bar A_{\pm}$ can be integrated out,
contributing in a trivial way to the vacuum functional. We could, therefore,
think of
${\cal L}_{g}$ as an effective theory, which represents
a scalar boson $\Phi$ in a gravitational background. Later we will
come back to this question.

\bigskip

To study the symmetries associated with ${\cal L}_{TOT}$,
in (\ref{132},\ref{133}), let us first note that
the original vectorial symmetry given by (\ref {20})
and (\ref {100}) is now hidden. Actually, since the metric is a function
only of the $\lambda$'s, it does not transform at all. The field
$\Phi$ is also invariant, which can be seen from (\ref {20}),
(\ref {100}) and (\ref{135}). Finally, from (\ref{130}), we see that
$\delta\bar A_{\pm}=0$.
Collecting all these facts, we see that (\ref{132}) is trivially invariant
under the local
vectorial symmetry.

Under a diffeomorphism

\begin{equation}
\label{136}
\delta x^\alpha=-\epsilon^\alpha\,,
\end{equation}

\noindent a symmetric tensorial density $G^{\alpha\beta}$ transforms as

\begin{equation}
\label{137}
\delta G^{\alpha\beta}=\partial_\gamma(G^{\alpha\beta}\epsilon^\gamma)
-G^{\gamma\alpha}\partial_\gamma\epsilon^{\beta}
-G^{\gamma\beta}\partial_\gamma\epsilon^{\alpha}\,.
 \end{equation}

\noindent  From (\ref{134}) and (\ref{137}),we derive after some algebraic
calculations
that under a diffeomorphism

\begin{eqnarray}
\label{138}
\delta\lambda_{++}&=&-\partial_+\epsilon^-+\lambda^2_{++}\partial_-\epsilon^+
+(\partial_+\epsilon^+-\partial_-\epsilon^-+\epsilon^+\partial_+
+\epsilon^-\partial_-)
\lambda_{++}\nonumber\\
\delta\lambda_{--}&=&-\partial_-\epsilon^++\lambda^2_{--}\partial_+\epsilon^-
+(\partial_-\epsilon^--\partial_+\epsilon^++\epsilon^+\partial_+
+\epsilon^-\partial_-)
\lambda_{--}\,.
\end{eqnarray}

\noindent  Since $\Phi$ transforms as a scalar under diffeomorphism,
i.e., $\delta\Phi=\epsilon^\alpha\partial_\alpha\Phi$,
${\cal L}_g$ in (\ref{133})  can be seen to be invariant modulo
total derivatives, while ${\cal L}_{\bar A}$ can be made
invariant once we choose

\begin{eqnarray}
\label{139}
\delta
\bar A_+=\epsilon^-\partial_-\bar A_+&+&\partial_+
\left(\epsilon^+\bar A_+\right)+
\bar A_-\partial_+\epsilon^-
-{1\over4}\partial_+\epsilon^-{\delta S\over{\delta \bar A_+}}\,,\nonumber\\
\delta
\bar A_-=\epsilon^+\partial_+\bar A_-&+&\partial_-
\left(\epsilon^-\bar A_-\right)+
\bar A_+\partial_-\epsilon^+
-{1\over4}\partial_-\epsilon^+{\delta S\over{\delta \bar A_-}}\,,
\end{eqnarray}

\noindent where

\begin{equation}
\label{138a}
{\delta S\over{\delta \bar A_\pm}}=2\left(\bar A_\mp+\lambda_{\pm\pm}\bar
A^\pm\right) \,.
\end{equation}

\noindent  We can see that the transformations for
$\bar A_\pm$ are just those of the vectors plus
terms that vanish on-shell.
\bigskip

After some lengthy algebraic calculations we can show
that the algebra of diffeomorphism closes on the field $\Phi$
as well as on the $\lambda$'s as

\begin{eqnarray}
\label{140}
\left[\delta_1\,,\delta_2 \right]\Phi&=&\delta_3\Phi\nonumber\\
\left[\delta_1 \,,\delta_2 \right]\lambda_{\pm\pm}&=&\delta_3\lambda_{\pm\pm}
\end{eqnarray}

\noindent where

\begin{equation}
\label{140a}
\epsilon^\alpha_3= -\epsilon_1^\beta\partial_\beta\epsilon_2^\alpha+
\epsilon_2^\beta\partial_\beta\epsilon_1^\alpha \,,
\end{equation}

\noindent while variations on $\bar A_{\pm}$ satisfy an open
algebra \cite{Gomis} of the form

\begin{eqnarray}
\label{140b}
\left[\delta_1,\delta_2\right]\bar A_+&=&\delta_3\bar A_+
+ V_{+-}{\delta S\over{\delta \bar A_-}}
\nonumber\\
\left[\delta_1,\delta_2\right]\bar A_-&=&\delta_3\bar A_-
+ V_{-+}{\delta S\over{\delta \bar A_+}}
\end{eqnarray}

\noindent where

\begin{equation}
\label{140c}
V_{+-}=-V_{-+}={1\over4}\left(\partial_+\epsilon_1^-\partial_-\epsilon
_2^+-\partial_+\epsilon_2^-\partial_-\epsilon_1^+\right)\,.
\end{equation}

\noindent  In other words, the algebra closes on-shell for the
$\bar A_\pm$ fields. This, in turn, implies that one can introduce
auxiliary fields to close the algebra off-shell.

\section{Truncations, dualities and generators}

So far we have shown that we can consider
${\cal L}_{g}$ as an effective Lagrangian after integrating out
the $\bar A_\pm$ fields
from the complete theory, which solders two initially
decoupled
Siegel bosons of opposite chiralities.

We would like to comment  that if we
restrict the diffeomorphism to just one sector,
say by requiring $\epsilon^+=0$,
we reproduce the original Siegel
symmetry for the sector described by the pair $\Phi\,,\lambda_{++}$
in the same way as it appears in (\ref{01},\ref{02}). However,
under this restriction,
$\lambda_{--}$ transforms in a non-trivial
way as

\begin{equation}
\label{c1}
\delta\lambda_{--}=\lambda^2_{--}\partial_+\epsilon^-+
\left(\partial_-\epsilon^-+\epsilon^-\partial_-\right)\lambda_{--}\,.
\end{equation}

\noindent  The original Siegel model, therefore, is
not a subgroup of the original diffeomorphism group but
it is only recovered if we also make a
further truncation, by imposing  that $\lambda_{--}=0$.
The existence of the residual symmetry (\ref {c1}) seems to be related
to a duality symmetry satisfied by ${\cal L}_g$ when the
metric is parametrized as in (\ref{134}).
Under the transformation

\begin{equation}
\label{c2}
\lambda_{\pm\pm}\rightarrow{1\over\lambda_{\mp\mp}}  \,,
\end{equation}

\noindent we see that (i) the set of variations (\ref{138}) is invariant
and (ii) that ${\cal L}_g$ goes into $-{\cal L}_g$, so that the equations of
motion are invariant. The duality
present in the equations of motion of the theory can also be seen
to work in a first order formalism. By introducing

\begin{equation}
\label{c3}
\Pi={\partial{\cal L}_g\over{\partial\left(\partial_+\Phi\right)}}
\end{equation}

\noindent as the momentum canonically conjugated to $\Phi$, we can rewrite
${\cal L}_g$ as

\begin{equation}
\label{c4}
{\cal L}_{F.O.}=\Pi\partial_+\Phi+{1\over2}\lambda_{++}T_{--}-
{1\over2}{1\over\lambda_{--}}T_{++} \,,
\end{equation}

\noindent where

\begin{equation}
\label{c5}
T_{\pm\pm}={1\over2}\left(\Pi\mp\partial_-\Phi\right)^2
\end{equation}

\noindent are the diffeomorphism generators and satisfy the Virassoro algebra.
We see from the above equations that the duality invariance of (\ref{c4})
is achieved not only with (\ref{c2}) but also with the change
$T_{\pm\pm}\rightarrow T_{\mp\mp}$, which is the same as having $x^-\rightarrow
-x^-$, but keeping $x^+$ unchanged. This is obviously related to the symmetry
under the interchange of the right and the left moving sectors of the theory.
In the next section we will come back to this point by writing in a geometric
manner the original left and right moving Siegel models.

\section{ Further geometrical considerations}

The Siegel model as well as the soldering can be described in a geometrical
manner also. Let us look at the Siegel model in one sector only and note that
we can write it also as

\begin{eqnarray}
\label{d1}
{\cal L}^{(+)}_0&=&\partial_+\varphi\partial_-\varphi+
\lambda_{++}(\partial_-\varphi)^2\nonumber\\
&=&{1\over2}\sqrt{-g}g^{\alpha\beta}\partial_\alpha\varphi\partial_\beta\varphi
\end{eqnarray}

\noindent where

\begin{equation}
\label{d2}
\sqrt{-g}g^{\alpha\beta}\,\equiv\, G^{\alpha\beta}_{(+)}
=\left(\begin{array}{cc}
0&1\\
1&{2\lambda_{++}}
\end{array}\right)\,.
\end{equation}

\noindent  From (\ref{d1}, \ref{d2}), we note that we can think of the Siegel
particle as propagating in a background gravitational field in a light cone
gauge
for which the invariant length has the form

\begin{equation}
\label{d3}
ds^2=2g_{+-}dx^+dx^-+g_{++}(dx^+)^2\,.
\end{equation}

\noindent  The Siegel invariance of Eq. (\ref{d1}) can be understood as
the residual one parameter coordinate invariance in this gauge, defined by

\begin{eqnarray}
\label{d4}
x^-&\rightarrow &x^--\epsilon^-(x^+,x^-)\nonumber\\
\delta\varphi&=&\epsilon^\alpha\partial_\alpha\varphi
\end{eqnarray}

\noindent where $\delta G^{\alpha\beta}_{(+)}$ is given by Eq. (\ref{137}) and
we
assume

\begin{equation}
\label{d5}
\epsilon^+=0  \,.
\end{equation}

\noindent Similarly, The Siegel Lagrangian in the other sector can also
be written as

\begin{eqnarray}
\label{d6}
{\cal L}^{(-)}_0&=&\partial_+\rho\partial_-\rho+
\lambda_{--}(\partial_-\rho)^2\nonumber\\
&=&{1\over2}\sqrt{-\tilde g}\tilde g^{\alpha\beta}
\partial_\alpha\varphi\partial_\beta\varphi
\end{eqnarray}

\noindent where

\begin{equation}
\label{d7}
\sqrt{-\tilde g}\tilde g^{\alpha\beta}\,\equiv
\,  G^{\alpha\beta}_{(-)}=\left(\begin{array}{cc}
{2\lambda_{--}}&1\\
1&0
\end{array}\right)\,.
\end{equation}

\noindent In other words, the Siegel particle in the other sector can
also be thought of
as moving in a background gravitational field with a light cone metric of the
form

\begin{equation}
\label{d8}
ds^2=2\tilde g_{+-}dx^+dx^-+\tilde g_{--}(dx^-)^2\,.
\end{equation}

\noindent  The Siegel invariance, in this sector, can again be thought of
as a one
parameter residual coordinate invariance of the form

\begin{equation}
\label{d9}
x^+\rightarrow x^+-\epsilon^+\left(x^-,x^+\right)\,.
\end{equation}

\noindent  The gauged Lagrangian in one sector, Eq. (\ref{70}),
cannot be written in a diffeomorphism invariant manner. Therefore,
gauging in one of the sectors breaks Siegel invariance. However,
let us note the following from Eq. (\ref{70}).

\begin{eqnarray}
\label{d10}
{\cal L}_2^{(+)}&=&(\partial_-\varphi)\left(\partial_+\varphi +\lambda_{++}
\partial_-\varphi\right)+2\left(\partial_+\varphi+\lambda_{++}\partial_-\varphi
\right)+\lambda_{++}A^2_-\nonumber\\
&=&-(\partial_+\varphi)\left(\partial_-\varphi+{1\over\lambda_{++}}\partial_+
\varphi\right)+\lambda_{++}\left(A_-+\partial_-\varphi
+{1\over\lambda_{++}}\partial_+\phi\right)^2\,.
\end{eqnarray}

\noindent  This shows that if we integrate out the $A_-$ field, the Siegel
theory changes
chirality with the identification

\begin{equation}
\label{d11}
\lambda_{--}={1\over \lambda_{++}}\,,
\end{equation}

\noindent that again, is related to the duality symmetry discussed in the
last section.

Finally, we note that the complete Lagrangian of Eq. (\ref{120}) can also be
written in form (\ref{132}), but we can rewrite ${\cal L}_A$ as

\begin{equation}
{\cal L}_A= {1\over2}M^{\alpha\beta}\bar A_\alpha \bar A_\beta \,,
\label{d12}
\end{equation}

\noindent where

\begin{eqnarray}
\label{d13}
M&=& \left(G_{(+)}+
G_{(-)}\right)=
2\left(\begin{array}{cc}{\lambda_{--}} &
1\\1&{\lambda_{++}}\end{array}\right)\nonumber\\
&=& -{\Delta\over2}\left(G-\sigma_1\right)\nonumber\\
\bar A&=&\left(\begin{array}{clcr}A_++{1\over2}\partial_+(\rho+\varphi)\\
A_-+{1\over2}\partial_-(\rho+\varphi)\end{array}\right)+
i\sqrt{2}\sigma_2  G\left(\begin{array}{c}\partial_+\Phi\\
\partial_-\Phi\end{array}\right)\,.
\end{eqnarray}

\noindent Here $\sigma_1$ and $\sigma_2$ represent the usual Pauli matrices.
We also note that the Lagrangian in Eq. (\ref{d12}), with the identifications
in Eq. (\ref{d13}), can also be written as

\begin{eqnarray}
\label{d14}
{\cal L}_{tot}&=& {1\over 2}
\sqrt{-g}g^{\alpha\beta}
\partial_\alpha\Phi\partial_\beta\Phi
-{1\over 4}
\sqrt{-g}g^{\alpha\beta}\Delta
 \bar A_\alpha\bar A_\beta+{\Delta\over4}
\bar A_+\bar A_-\,,\nonumber\\
&=& {1\over 2}
\sqrt{-g}g^{\alpha\beta}
\partial_\alpha\Phi\partial_\beta\Phi
-\sqrt{-g}g_{\alpha\beta}
 \hat{A}_\alpha\hat{A}_\beta+\hat{A}_+\hat{A}_-\,.
\end{eqnarray}

\noindent where we have defined $\hat A_\alpha={\sqrt{\Delta}\over2}\bar
A_\alpha$.
There are several things to note here. First, the combination of the
scalar fields, $\varphi+\rho$, has gone into the redefinition of the vector
field. From the structure of the Lagrangian, it is obvious that the original
gauge transformations have disappeared completely, as we
saw in the beginning of the
last section  3. The diffeomorphism invariance of the theory is almost
obvious. The presence of the last term in Eq. (\ref{d14}) suggests that
$\hat{A}$ cannot transform really like a vector under a coordinate invariance.
In fact, ${\cal L}_{TOT}$ can be checked to be invariant under the set of
transformations (\ref{137}), the already quoted diffeomorphism transformation
for $\Phi$ as well as

\begin{equation}
\label{d15}
\delta\hat{A}_\alpha=\hat{A}_\beta\partial_\alpha\epsilon^\beta+
\epsilon^\beta\partial_\beta\hat{A}_\alpha+
{1\over2}M^{-1}_{\alpha\beta}\left(\begin{array}{cl}
\hat{A}_+\partial_-\epsilon^+ \\
\hat{A}_-\partial_+\epsilon^-\end{array}\right)^\beta\,.
\end{equation}

\noindent  The algebra, of course, trivially closes on the $\Phi$ and
$G^{\alpha\beta}$ variables. For $\hat{A}_\alpha$, however, we can show that

\begin{equation}\label{d16}
\left[\delta_1,\delta_2\right]\hat A_\alpha=
\delta_3\hat{A}_\alpha + {1\over 2\Delta^2}\left((i\sigma_2)M\,\hat
A\right)_\alpha
\left(\partial_+\epsilon^-_2\partial_-\epsilon^+_1-
 \partial_+\epsilon^-_1\partial_-\epsilon^+_2\right)\,.
\end{equation}

\noindent As expected, the algebra of the variations
closes on-shell for the gauge fields.
\section{Conclusions}

In this work we have shown how to solder two initially decoupled Siegel
bosons of different chiralities. This has been done through the implementation
of a vector gauge symmetry, which has forced the two bosons to belong to
the same multiplet. The complete theory so obtained presents
full diffeomorphism
invariance and can be represented in a geometric manner. We have shown, as
expected, that the diffeomorphism algebra closes on the fields appearing
in the theory. The way we have parametrized the metric has made explicit
that the naive sum of two Siegel metrics is not the metric of a full
diffeomorphism invariant theory. In this sense we could reveal the
relashionship between Siegel symmetry and diffeomorphism.  We have also
discovered a surprising duality in the model, which is related to
the symmetry under the exchange between the left and right movers.\vspace
{0.1cm}

\bigskip

\noindent {\bf ACKNOWLEDGEMENTS}  This work has been supported in part by
U.S. Department of Energy, grant No DE-FG-02-91ER 40685, and by CNPq,
Brazilian research agency, Brasilia, Brazil.

\vfill\eject

\end{document}